\documentclass{aa}
\usepackage{graphicx}
\usepackage{times,psfig}

\def\ltsima{$\; \buildrel < \over \sim \;$}
\def\lsim{\lower.5ex\hbox{\ltsima}}

\def\gtsima{$\; \buildrel > \over \sim \;$}
\def\gsim{\lower.5ex\hbox{\gtsima}}

\newcommand{\be}{\begin{equation}}
\newcommand{\en}{\end{equation}}

\begin{document}
%\received{~~} \accepted{~~}
%\journalid{}{}
%\articleid{}{}

\title{Are \textit{Swift} gamma--ray bursts consistent with the Ghirlanda relation?}

\authorrunning{S. Campana et al.}
\titlerunning{Are \textit{Swift} GRBs consistent with the
Ghirlanda relation?}

%\subtitle

\author{S.~Campana\inst{1} \and
C. Guidorzi\inst{2,1} \and G. Tagliaferri\inst{1} \and 
G. Chincarini\inst{2,1} \and  A. Moretti\inst{1} \and 
D. Rizzuto\inst{2,1} \and P. Romano\inst{1,2}} 
 
\institute{
{INAF-Osservatorio Astronomico di Brera, via E. Bianchi 46,
I--23807 Merate (Lc), Italy}
\and
{Universit\`a degli studi di Milano-Bicocca, Dipartimento di
Fisica, piazza delle Scienze 3, I--20126 Milano, Italy}
}

\offprints{campana@merate.mi.astro.it}

\abstract{
A few tight correlations linking several
properties of gamma--ray bursts (GRBs), namely the spectral peak energy, the
total radiated energy, and the afterglow break time, have been discovered with
pre-\textit{Swift} GRBs. They were used to constrain the cosmological parameters,
together with type-Ia supernovae. However, the tightness of these correlations
is a challenge to GRB models. We explore the effect of adding 
\textit{Swift} bursts to the Ghirlanda and Liang-Zhang relations. Although they 
are both still valid, they become somewhat weakened mostly due to
the presence of significant outliers, which otherwise are apparently normal
GRBs so difficult to distinguish. The increased dispersion of
the relations makes them less reliable for purposes of precision cosmology.


}

\maketitle

\section{Introduction}

Gamma--ray bursts (GRBs) release a huge amount of energy in a short amount of
time, making them the brightest sources in the gamma--ray sky. Following these
explosions, a residual emission, the afterglow, is observed at X--ray and,
often, at UV-optical-IR wavelengths (see van Paradijs  et al. 2000 and
Zhang \& M\'esz\'aros 2004 for reviews). Given the involved energies, GRBs
represent the brightest objects in the Universe and can therefore be detectable
up to very high redshifts. In contrast to type-Ia supernovae, which are good
cosmological rulers up to redshift $z\sim 1$ (e.g. Perlmutter et al. 1999;
Riess et al. 2004), it has been speculated that GRBs are promising standard candles 
for a high redshift universe, and attempts to use GRB properties to indicate 
distances have been pursued in the past.

Based on samples of varying size, the isotropic-equivalent burst
luminosities were found to be correlated with the temporal variability (smooth
bursts are intrinsically less luminous; Fenimore \& Ramirez-Ruiz 2000; Reichart
et al. 2001; Guidorzi et al. 2005) and anti-correlated with the spectral time
lags (the times elapsing between the light curve at higher and lower energies;
Norris et al. 2000).
There were also encouraging results when considering spectral properties. 
Amati et al. (2002) found a correlation between the rest-frame peak energy 
$E'_{\rm p} = (1+z)E_{\rm p}$
and the isotropic-equivalent energy released during the prompt phase $E_{\rm
iso}$. This correlation, initially based on 10 GRBs with known redshift,
was later updated and confirmed using a sample of 41 GRBs (Amati 2006, 2007; 
see however Band \& Preece 2005; Nakar \& Piran 2005). The Amati relation has
been used for a number of statistical studies to derive the redshift of GRBs
with known peak energy and fluence (e.g. Atteia 2003; Liang et al. 2004). 
However, the relatively large scatter results in poorly constrained redshifts 
(especially for high values). This indicates
that  the Amati relation is not tight enough for cosmological studies (e.g. Li
2006). To date, there are two known outliers of the Amati relation,
namely GRB\,980425 and GRB\,031203. These bursts are peculiar in several
respects, since they are very close ($z \le 0.1$), underluminous with respect to
most other events, and they have peculiar afterglows (e.g. Woosely \& Bloom 2006).
Driven by the observation that XRF060218 showed a strong spectral
evolution during its prompt emission phase (Campana et al. 2006), and that the
mean peak energy is consistent with the Amati relation (Amati et al. 2007),
Ghisellini et al. (2006) speculate that these two outliers may also be
consistent with the Amati relation, if a strong spectral evolution took place.

A significant step forward was performed by Ghirlanda et al. (2004a). Since
GRBs are believed to originate inside jets (with half-opening angle
$\vartheta$), they considered the collimation-corrected energy $E_\gamma =
(1-\cos\vartheta) E_{\rm iso}$ instead of the isotropic-equivalent value.
Remarkably, based on a set of 15 GRBs (now expanded to 18; Nava et al. 2006),
$E_\gamma$ is much more tightly correlated with $E'_{\rm p}$ than is $E_{\rm
iso}$ (see also Friedman \& Bloom 2005). With such tightness, Ghirlanda et
al. (2004b) combined 15 GRBs with data from type-Ia supernovae and were able to
constrain the cosmological parameters (see also Xu et al. 2005 and Schaefer
2007 and references therein).
%$\Omega_{\rm m} = 0.37\pm0.10$ and $\Omega_\Lambda = 0.87\pm0.23$. 
To infer the jet collimation degree, $\vartheta$ is computed from the time at
which the afterglow light curve steepens (Rhoads et al. 1999; Sari et al.
1999; Chevalier \& Li 2000). There are two possible cases, depending on the
density profile of the medium surrounding the GRB source, which is usually
assumed either homogeneous (interstellar medium, ISM) or wind-shaped ($n = A
\times r^{-2}$, wind). In particular, we have (Sari et al. 1999; Chevalier \&
Li 2000)
\begin{equation}\label{eq:theta}
  \vartheta = \left\{\begin{array}{ll}
    \displaystyle 0.161\left(\frac{t_{\rm b,d}}{1+z}\right)^{3/8}
    \left(\frac{n_0\,\eta_\gamma}{E_{\rm iso,52}}\right)^{1/8}~\mathrm{rad} &
      \mathrm{ISM} \\
    \displaystyle 0.202\left(\frac{t_{\rm b,d}}{1+z}\right)^{1/4}
    \left(\frac{A_*\,\eta_\gamma}{E_{\rm iso,52}}\right)^{1/4}~\mathrm{rad} &
      \mathrm{wind} \\
  \end{array}\right.
\end{equation}
where $t_{\rm b,d}$ is the break time in days, $\eta_\gamma$ the radiation
efficiency, $E_{\rm iso,52} = E_{\rm iso}/(10^{52}~\mathrm{erg})$, $n_0$ is the
ambient particle density in cm$^{-3}$, and $A_* = 3 \times
10^{35}~{\rm cm}^{-1}$ is the wind constant.

The Ghirlanda relation is tight in both cases, with a logarithmic dispersion of
0.10 dex (0.08 dex) in the ISM (wind) case; in comparison, the Amati relation
has a dispersion of 0.4 dex. It is not straightforward to compare GRB\,980425
and GRB\,031203 with the Ghirlanda relations, since their collimation degree is
unknown, but it looks likely that these events are outliers for them as well.
We also note that the Ghirlanda relation is intimately connected with the
jet break interpretation. The standard jet break theory predicts 
a variation in the decay index of $\Delta\alpha\sim 1$ and a decay index after the
break $\alpha_2\gsim 2$. These predictions are loosely satisfied by the
pre-\textit{Swift}
sample. By taking the full analysis of Zeh,  Klose \& Kann (2006), we find
that only 3 GRBs of the 13 pre-\textit{Swift} bursts available (5 GRBs do not have 
enough data) have a $\Delta\alpha\sim 1$ at $1\,\sigma$ level. Allowing a
$3\,\sigma$ level we have 9 GRBs consistent with this constraint, leaving out 4
GRBs that do not satisfy these relations. In addition, 2 bursts definitely have
($3\,\sigma$) a decay which is shallower than 2. Several different explanations can
address this issue as the uncertainty of sideways expansion physics (e.g. Panaitescu et
al. 2005), different electron energy distributions ($p$ value) for different
GRBs, change in the microphysical parameters ($\epsilon$), new energy injections, 
changing external medium, etc.

The need to estimate the jet angle through Eq.~(\ref{eq:theta}) somehow makes the
Ghirlanda relations model-dependent. Recently, a lot of discussion has
arisen concerning the interpretation of breaks in the afterglow light curves of
GRBs discovered by \textit{Swift} (e.g. Panaitescu et al. 2006; Sato et
al. 2007; Willingale et al. 2007; Zhang 2006). For example, Panaitescu et al. (2006)
present six optical/X-ray afterglows whose light curves clearly show a
chromatic behavior, which cannot be due to a geometric effect. Liang \& Zhang
(2005) have previously pointed out that a purely phenomenological correlation
(i.e. without any assumption on the nature of the breaks)
exists between $E_{\rm iso}$, $t'_{\rm b} = t_{\rm b}/(1+z)$, and $E'_{\rm p}$,
in the form $E_{\rm iso} \propto {E'_{\rm p}}^x {t'_{\rm b}}^y$. It should be
noted that Liang \& Zhang (2006) argue that the breaks in optical and X--ray bands
may have different physical origins. They explicitly took $t_{\rm b}$ from
\textit{optical} light curves (actually, this was also the case for most of
the Ghirlanda sample). 
Nava et al. (2006) point out that the Ghirlanda and Liang-Zhang
relations are mutually consistent, provided that $y=-1$ (as their data
support). 

During the past two years, the GRB field has been revolutionized by
\textit{Swift}. More than 200 long-duration GRBs have been discovered and
a secure redshift has been obtained for more than 50 (up to February 2006). 
\textit{Swift} GRBs with known redshift are at a higher mean
redshift than before (Jakobsson et al. 2006). GRBs at larger redshift
might have different characteristics maybe related to metallicity's impact
on these correlations.
Obviously it is very relevant to see how these new
GRBs behave with respect to the Ghirlanda and Liang-Zhang relations. To this
aim, three parameters are needed for each event, namely the redshift, the
spectral peak energy, and the afterglow break time. The burst alert
telescope (BAT) onboard \textit{Swift}, which detects the bursts, covers a
limited spectral energy range (15--350 keV), so that $E_{\rm p}$ can only be
determined in very few cases. 
However, some GRBs discovered by \textit{Swift} can also be observed by other
instruments, in particular by the \textit{Wind}-Konus experiment, which has a
useful energy range of 18--1160/21--1360 keV, or by HETE-2,
\textit{Suzaku}, and RHESSI. In this paper, we present a
systematic search for all \textit{Swift} bursts, which have the measurement of the
three parameters needed for comparing them with the Ghirlanda and Liang-Zhang
relations.

When this paper was submitted, we became aware of a similar
paper (Ghirlanda et al. 2007). We have not carried out a detailed comparison
between the two works but highlight a few differences in the following sections.

\begin{table*}

\caption{Sample of \textit{Swift} GRBs with known $z$, $E_{\rm p}$, and $t_{\rm b}$.\label{tbl-2}}
\begin{center}\begin{tabular}{l@{$\;\;$}c@{$\;\;$}c@{$\;\;$}c@{$\;\;$}c@{$\;\;$}c@{$\;\;$}c@{$\;\;$}c@{$\;\;$}c@{$\;\;$}c@{$\;\;$}c} \hline\hline
GRB     & $z$   & $E'_{\rm p}$ & $t_{\rm b}$       & $\log E_{\rm iso}$ & $\vartheta^{\rm (h)}$ &$\vartheta^{\rm (w)}$ & $\log E_\gamma^{\rm (h)}$  & $\log E_\gamma^{\rm (w)}$ & Pred. $t_{\rm b}^{\rm (h)}$ & Pred. $t_{\rm b}^{\rm (w)}$ \\
        &       & (keV)        & (d)          & (erg)          & (deg)       & (deg)       & (erg)          & (erg)          & (d)            & (d)               \\
(1)     & (2)   & (3)          & (4)          & (5)            & (6)         & (7)         & (8)            & (9)            & (10)                & (11)                \\ \hline
050525A & 0.606 & 127$\pm$10   & 0.3$\pm$0.1   & 52.53$\pm$0.03 & 4.0$\pm$0.7 & 3.7$\pm$1.0 & 49.91$\pm$0.15    & 49.86$\pm$0.16 & 0.5$^{+0.1}_{-0.1}$ & 0.5$^{+0.4}_{-0.2}$ \\
050820A & 2.612 & 1325$\pm$270 & 15$\pm$8      & 53.92$\pm$0.09 & 8.5$\pm$2.0 & 3.7$\pm$1.1 & 51.96$\pm$0.17    & 51.23$\pm$0.16 & 4.2$^{+2.9}_{-1.7}$ & 4.5$^{+4.7}_{-2.3}$ \\
050922C & 2.198 & 415$\pm$111  & 0.11$\pm$0.03 & 52.79$\pm$0.17 & 2.0$\pm$0.3 & 2.1$\pm$0.6 & 49.55$\pm$0.23    & 49.62$\pm$0.27 & 5.4$^{+4.9}_{-2.5}$ & 5.7$^{+7.0}_{-3.2}$ \\
051109A & 2.346 & 539$\pm$381  & 0.60$\pm$0.10 & 52.70$\pm$0.22 & 3.7$\pm$0.4 & 3.4$\pm$0.6 & 50.02$\pm$0.26    & 49.94$\pm$0.32 & 11$^{+36}_{-9}$     & 12$^{+43}_{-9}$     \\
060124  & 2.297 & 636$\pm$162  & 1.13$\pm$0.09 & 53.62$\pm$0.06 & 3.6$\pm$0.5 & 2.3$\pm$0.6 & 50.93$\pm$0.09    & 50.54$\pm$0.15 & 1.8$^{+1.4}_{-0.8}$ & 2.0$^{+2.1}_{-1.0}$ \\
060206  & 4.045 & 381$\pm$98   & 0.6$\pm$0.2   & 52.76$\pm$0.08 & 3.1$\pm$0.6 & 2.9$\pm$0.8 & 49.93$\pm$0.17    & 49.88$\pm$0.19 & 7.6$^{+5.7}_{-3.2}$ & 8.0$^{+8.7}_{-4.2}$ \\
060526  & 3.21  & 105$\pm$21   & 2.77$\pm$0.30 & 52.41$\pm$0.05 & 5.5$\pm$0.4 & 3.9$\pm$0.5 & 50.23$\pm$0.08    & 50.07$\pm$0.11 & 0.3$^{+0.2}_{-0.1}$ & 0.3$^{+0.3}_{-0.1}$ \\
060614  & 0.125 & 55$\pm$45    & 1.39$\pm$0.04 & 51.40$\pm$0.22 & 11.1$\pm$0.9& 11.5$\pm$1.8& 49.67$\pm$0.25    & 49.70$\pm$0.31 & 0.3$^{+0.9}_{-0.2}$ & 0.3$^{+1.0}_{-0.2}$ \\
\hline
050318  & 1.44  & 115$\pm$25   & $>$0.26       & 52.41$\pm$0.03 & $>$3.3 (0.7) & $>$3.5 (0.6) & $>$49.63 (0.22) & $>$49.68 (0.20) & 0.8$^{+0.5}_{-0.3}$ & 0.9$^{+0.8}_{-0.4}$ \\
050401  & 2.90  & 501$\pm$53   & $>$13         & 53.61$\pm$0.08 & $>$8.5 (2.0) & $>$4.1 (1.2) & $>$51.66 (0.16) & $>$51.03 (0.15) & 1.4$^{+0.6}_{-0.4}$ & 1.5$^{+1.2}_{-0.7}$ \\
050416A & 0.653 & 25.1$\pm$4.2 & $>$4.7        & 51.08$\pm$0.07 & $>$16.7 (3.8)& $>$17.0 (5.0)& $>$49.70 (0.15) & $>$49.72 (0.15) & 0.6$^{+0.4}_{-0.2}$ & 0.6$^{+0.6}_{-0.3}$ \\
050603  & 2.821 & 1333$\pm$107 & $>$2.5        & 53.85$\pm$0.03 & $>$4.3 (1.0) & $>$2.4 (0.7) & $>$51.30 (0.22) & $>$50.79 (0.20) & 5.3$^{+1.9}_{-1.4}$ & 5.7$^{+4.3}_{-2.4}$ \\
060418  & 1.489 & 572$\pm$114  & $>$5          & 52.95$\pm$0.05 & $>$8.5 (2.0) & $>$5.3 (1.6) & $>$51.00 (0.23) & $>$50.59 (0.20) & 5.3$^{+3.7}_{-1.9}$ & 5.6$^{+5.2}_{-2.7}$ \\
060927  & 5.6   & 473$\pm$116  & $>$0.16       & 52.92$\pm$0.09 & $>$1.6 (0.4) & $>$1.8 (0.5) & $>$49.54 (0.25) & $>$49.61 (0.23) & 10$^{+8}_{-4}$      & 11$^{+12}_{-6}$     \\
061007  & 1.261 & 902$\pm$43   & $>$1.74       & 54.00$\pm$0.04 & $>$5.3 (1.2) & $>$2.6 (0.8) & $>$51.47 (0.15) & $>$50.90 (0.14) & 1.0$^{+0.3}_{-0.2}$ & 1.1$^{+0.8}_{-0.5}$ \\
061121  & 1.314 & 1288$\pm$153 & $>$3.5        & 53.36$\pm$0.04 & $>$6.8 (1.6) & $>$3.9 (1.1) & $>$51.21 (0.23) & $>$50.73 (0.20) & 1.8$^{+0.7}_{-0.5}$ & 1.9$^{+1.5}_{-0.8}$ \\ \hline
\end{tabular}
\end{center}

{(1) -- GRB name. (2) -- redshift. (3) -- Rest-frame spectrum
peak energy. (4) -- Observed break time {in the optical band and in the
observer frame}. (5) --
Isotropic-equivalent bolometric energy. (6),(7) -- Jet opening angle computed
for  homogeneous and wind environments. (8),(9) -- Beaming-corrected energy
for the homogeneous and wind cases. (10),(11) -- Predicted break time assuming
the Ghirlanda relation as computed by Nava et al. (2006) after correcting for
a small typo in the published relation (Nava, private communication).} 

GRB060614 is not used in the analysis described in the text, but it is
reported here for completeness.
\end{table*}

\section{Sample selection}

We collected all long-duration GRBs detected by \textit{Swift} that have a
secure (spectroscopic) redshift $z$ and a published peak energy $E_{\rm p}$. We
found 19 events satisfying this criterium. We thus inspected their afterglow light
curve looking for breaks. Due to the unclear multi-wavelength behavior of these
breaks, we did not look at X--ray or radio data. This is mainly dictated
by the fact that, when the Ghirlanda relation (and similar works) was found, no
simultaneous optical and X--ray light curve existed. This choice (sticking to
optical light curves) also guarantees an unbiased extension of the Ghirlanda
relation. In any case, when describing the GRBs below, we outline when
differences from the standard jet model are occurring. 
We end up with eight GRBs that have a reliable measurement of $t_{\rm b}$ 
(GRB\,060614 is not used however). For eight more, we can only provide
a lower limit to this parameter. For the other three there is not published
information (yet) to constrain it. To compute the beaming-corrected energy, we first
need the isotropic-equivalent value. Recently, Amati (2006, 2007) listed the values
of $E_{\rm iso}$ and $E'_{\rm p}$ for 18 \textit{Swift} bursts, and we adopted his
values whenever available. In other cases, we resorted to values published in the
GCNs\footnote{See \texttt{http://gcn.gsfc.nasa.gov}\,.} (see below), computing
a bolometric correction when necessary. Following Ghirlanda et al. (2004), 
for all bursts we assumed an external density $n_0 = 3$~cm$^{-3}$
and a wind parameter $A_* = 1$ (for both quantities we also introduced an error
of $\Delta{\log_{10} n(A_*)}=0.5$). When dealing with lower limits in the break
times, we conservatively assumed a $50\%$ error on that limit. In addition, we 
consider the errors in logarithmic space by taking the left/right error closer 
to the best fit line (this has been done also for GRBs in the Nava's sample). 
Below we summarize their properties and all the values are 
reported in Table~\ref{tbl-2}. 

$\bullet$ GRB\,050318 was observed by \textit{Swift} and the BAT was able to determine
the peak energy (Perri et al. 2005). From the UVOT data, one can determine a
lower limit on the time of a possible break in the optical light curve of 0.26
d (Still et al. 2005). In the X--ray, a break around 0.2 d is also seen (Perri
et al. 2005). 

$\bullet$ GRB\,050401 was observed both by \textit{Swift}-BAT and 
\textit{Wind}-Konus. We take for $E_{\rm p}$ the average of the two values
presented by Golenetskii et al. (2005a). In the optical, the well-monitored
light curve (Rykoff et al. 2005; Watson et al. 2006) showed a slow,
uninterrupted decay up to 13~d after the burst. In the X--ray band, the
afterglow was observed up to 11.6~d after the trigger (De Pasquale et al.
2005), revealing a break at $0.06 \pm 0.01$~d (a value inconsistent with the
Ghirlanda relation; De Pasquale et al. 2005).
Ghirlanda et al. (2007) indicate a possible break in the optical light
curve at $1.5\pm0.5$ d. We note that a simple power-law fit is able to
reproduce all the observed optical data ($\chi^2_{\rm red}=1.3$ with 28 degrees of
freedom, with a null hypothesis probability of $17\%$). We also tried to fit
the same data with a smoothly-joined power-law without finding any statistical
evidence of a break.

$\bullet$ XRF\,050416A has a well-defined X--ray light
curve (Mangano et al. 2007a), showing the common steep-flat-steep behavior
(Nousek et al. 2006; O'Brien et al. 2006). No additional breaks are visible
between 0.017 and $\sim 42$ d after the GRB (Sakamoto et al. 2006; Mangano et
al. 2007a). Thanks to its softness it was possible to compute the peak energy
with BAT (Sakamoto et al. 2006). In the optical there are also no signs of a
break in the light curve up to at least $\sim 4.7$ d (see the $I$-band light
curve presented by Holland et al. 2007; see also Soderberg et al. 2007). 
Ghirlanda et al. (2007) fitted the optical light curve, including a
XRF\,060218-like light curve (Campana et al. 2006). In their fit they find a
break at $1.0\pm0.7$ d. We note that the SN presence is suggested by
Soderberg et al. (2006) and they do not require the presence of a break 
in the light curve in their fit with the SN modeled with a SN1998bw
template. 

$\bullet$ GRB\,050525A was a bright burst observed by \textit{Swift} and
\textit{Wind}-Konus (Blustin et al. 2005; Golenetskii et al. 2005b). It has
already been included in the sample of Nava et al. (2006). Several groups
(Blustin et al. 2005; Mirabal et al. 2005; Della Valle et al. 2006a) report a
break in the optical about 0.3~d after the GRB. We adopt here $t_{\rm b} = 0.3
\pm 0.1$~d, which is the average of the values found by these authors (see also
Nava et al. 2006). In the X--ray band, a break was also detected, at about
the same time as the optical one, suggesting a jet origin for this afterglow 
(Blustin et al. 2005). Sato et al. (2007), however, question this 
interpretation, since the post-break decay indices were different in the 
optical and X--ray bands, contrary to model expectations, and the decay
index after the break is too shallow in the X--ray and optical bands. 

$\bullet$ The X--ray light curve of the GRB\,050603 afterglow was a simple
power law between 0.4 and 7 d after the GRB (Grupe et al. 2006). The peak
energy was estimated through \textit{Wind}-Konus observations (Golenetskii et
al. 2005c). UVOT
observed the optical afterglow as well, showing no breaks within the interval
0.4--2.5 d (Grupe et al. 2006). A slope $\alpha=1.8\pm0.2$ was found 
by UVOT, consistent with the XRT one. Given the relatively steep value, it is
unclear whether a break should have appeared before or after the \textit{Swift}
observations. The Ghirlanda relation would predict a jet break time around
4--7 d, so we consider $t_{\rm b} > 2.5$ d.

$\bullet$ GRB\,050820A was a bright burst with large $E_{\rm iso}$ and $E'_{\rm
p}$ (\textit{Wind}-Konus, Cenko et al. 2006). The X--ray light curve of
GRB\,050820A shows the usual 
triple power-law behavior. A lower limit to any further break in the X--rays
can be set to $t_{\rm b} \gsim 17$ d. The optical light curve tracks the
X--ray one, but thanks to late-time HST images a break is detected at very late
times, and its timing is difficult to constrain. Using the data from Cenko et al.
(2006) and allowing for a late-time temporal slope in the range $1.5 < \alpha <
3$, one gets $7~{\rm d} < t_{\rm b} < 23$~d. This is consistent with the
value $t_{\rm b} = 18 \pm 2$~d found by Cenko et al. (2006) which imposed
$\alpha = 2.34 \pm 0.06$.

$\bullet$ The X--ray and optical light curves of GRB\,050922C have been presented by
Panaitescu et al. (2006). Both the X--ray and optical light curves show a
break, but at different times. A multi-band fit of the UVOT optical data
provides a break time at $2.7 \pm 0.7$ hr (Li et al. 2005; see also Andreev et
al. 2005). The currently published data show no further breaks up to $t\sim 
1.2$~d. The peak energy was estimated by HETE-2 observations (Crew et al.
2005), even if \textit{Wind}-Konus has failed to reveal an exponential cut-off
(Golenetskii et al. 2005d). 

$\bullet$ GRB\,051109A was discovered by \textit{Swift} and also observed by
\textit{Wind}-Konus (Golenetskii et al. 2005e). The sparse optical data
constrain the break time to be $t_{\rm b} > 0.64$~d (Pavlenko et al. 2005).
More recently Yost et al. (2007), suggested the presence of a break occurring at 
$0.6\pm0.1$ d in the optical light curve. This break however does not coincide
with an X--ray break and is interpreted as a cooling break. 

$\bullet$ GRB\,060124 is an astonishing burst. \textit{Swift} was
triggered by the precursor, allowing to study in detail the prompt phase with 
CCD X--ray spectroscopy for the first time (Romano et al. 2006). In the X--ray
band, a break in the light curve occurs at $1.21 \pm 0.17$ d. At variance with
many other \textit{Swift} GRBs, a break in the optical light curve was
observed nearly simultaneously with the X--ray break at $1.13 \pm 0.09$ d
(D. A. Kann, priv. comm.; see also Curran et al. 2007). The spectral peak
energy was computed by simultaneously fitting \textit{Wind}-Konus and BAT data
(Romano et al. 2006); note that the value reported by HETE-2 (Lamb et
al. 2006) is slightly higher, but still consistent with the adopted value. 
Also in this case, the X--ray decay index after the break is too shallow
for a jet break.

$\bullet$ The late optical and X--ray light curves of GRB\,060206 have been published
by Monfardini et al. (2006) and Stanek et al. (2006). In both bands the light
curve is complex, but a clear break in the optical curve can be detected at
$t_{\rm b} = 0.6$~d. The light curve then goes on uninterrupted until at least
2.3~d (Stanek et al. 2006; Monfardini et al. 2006). We take the $E_{\rm p}$
and $E_{\rm iso}$ from the value measured by \textit{Swift}-BAT (Palmer et al.
2006).

$\bullet$ The X--ray light curve of GRB\,060418 shows several bright flares. On
the other hand, the optical/NIR light curve is smooth (Molinari et al. 2007),
with no breaks between $\sim 150$~s and $\sim 5$~d. The $E_{\rm p}$ and
$E_{\rm iso}$ values have been taken from the Konus data (Golenetskii et al. 2006a).

$\bullet$ GRB\,060526 was discovered by \textit{Swift}, showing two strong
flares just after the main event. A simultaneous break in the optical and
X-ray light curves is observed at $\sim 2.8$ d (Dai et al. 2007). The
bolometric source flux and the peak energy were taken by Schaefer (2007).

$\bullet$ GRB\,060614 was a remarkable low-redshift burst that did not show
supernova signatures down to very deep limits (Della Valle et al. 2006b;  Fynbo
et al. 2006a; Gal-Yam et al. 2006). The early-time light curve was  complex,
but, remarkably, VLT and \textit{Swift}-XRT data show an achromatic break at
$\sim 1.3$~d (Mangano et al. 2007b). The peak energy
of this event is constrained by the \textit{Wind}-Konus data (Golenetskii et
al. 2006b), which present a time-resolved analysis. Amati et al. (2007) have
estimated the average $E_{\rm p}$ to lie in the range 10--100~keV range. Since
the BAT spectrum shows no deviation from a power-law (Gehrels et al. 2006) in
the range 15--150 keV, we further assumed $E_{\rm p} < 50$~keV. We took the
value of $E_{\rm iso}$ from Amati et al. (2007). Given the lack of a bright
supernova, however, there is an ongoing debate about whether this GRB is related to a
massive progenitor or to a merging of two compact objects (Gehrels et al.
2006; Zhang et al. 2006). Since the Amati and Ghirlanda relations hold for
long GRBs, we do not consider this burst in the following.  

$\bullet$ GRB\,060927 was a high-redshift burst at $z = 5.6$ (Fynbo et al.
2006b). The \textit{Swift} observations allow $E_{\rm p}$ to be constrained inside
the BAT range. We computed the rest-frame 1--10000 keV fluence by employing the
spectral parameters provided by Stamatikos et al. (2006) and found $E_{\rm iso}
= (8.4 \pm 1.6) \times 10^{52}$~erg. The optical data only imply a limit
$t_{\rm b} > 0.1$~d (e.g. Antoniuk et al. 2006). The X--ray light curve shows
a break at $t = 0.05$~d (Troja et al. 2006).

$\bullet$ The X--ray and optical light curves of GRB\,061007 consist of a
single power law starting soon after the burst. Current limits show $t_{\rm b}
> 11$ and 1.7 d after the GRB in the X--ray and optical bands, respectively
(Mundell et al. 2007; Schady et al. 2007). Schady et al. (2007) suggest that
a break might have occurred before the start of the XRT and UVOT observations
(72 s after the burst). However, ROTSE observations starting 26 s after the
burst trigger imply a rising flux between the two measurements, making this
hypothesis unlikely. Thus we chose $t_{\rm b} > 1.7$~d. We took $E_{\rm p}$
and $E_{\rm iso}$ from Golenetskii et al. (2006c), whose results are consistent
with the RHESSI values (Wigger et al. 2006). We note that \textit{Suzaku}-WAM
found a larger $E_{\rm p}$ (Yamaoka et al. 2006).

$\bullet$ GRB\,061121 showed a break at X--ray wavelengths at 3~ks after
the GRB (Page et al. 2006). In the optical band, there is a possible
rebrightening at $\sim 3$~ks, and the decay is then regular up to $\sim 3$~d
(Halpern \& Armstrong 2006; Efimov et al. 2006). Determining the location of a
break is complicated by the presence of a bright host galaxy (Malesani et al.
2006; Cobb 2006). We took the average of the $E_{\rm p}$ values as found by
\textit{Wind}-Konus (Golenetskii et al. 2006d) and RHESSI (Bellm et al. 2006).

$\bullet$ There are three more GRBs with measured $z$ and $E_{\rm p}$,
namely GRB\,060115, GRB\,060218, and GRB\,060707. For these bursts, it is
however impossible to measure a break time, due to the small amount of available 
data or to the presence of SN\,2006aj in the case of GRB\,060218 (e.g. Campana
et al. 2006; Pian et al. 2006). We encourage observers with available data to
provide a measurement for the break time of these bursts.

The values that we adopted for the optical break times are strictly compliant
to the selection made by Ghirlanda et al. (2004a). The assumed error budget is
conservative. By taking the limits from the \textit{Swift} X--ray light curves
(which often extend much further in time), we would find even tighter limits, as
recently shown by Sato et al. (2007) for a subsample of our GRBs.

\section{Analysis and results}

\begin{figure*}[htbp]
\begin{center}
\psfig{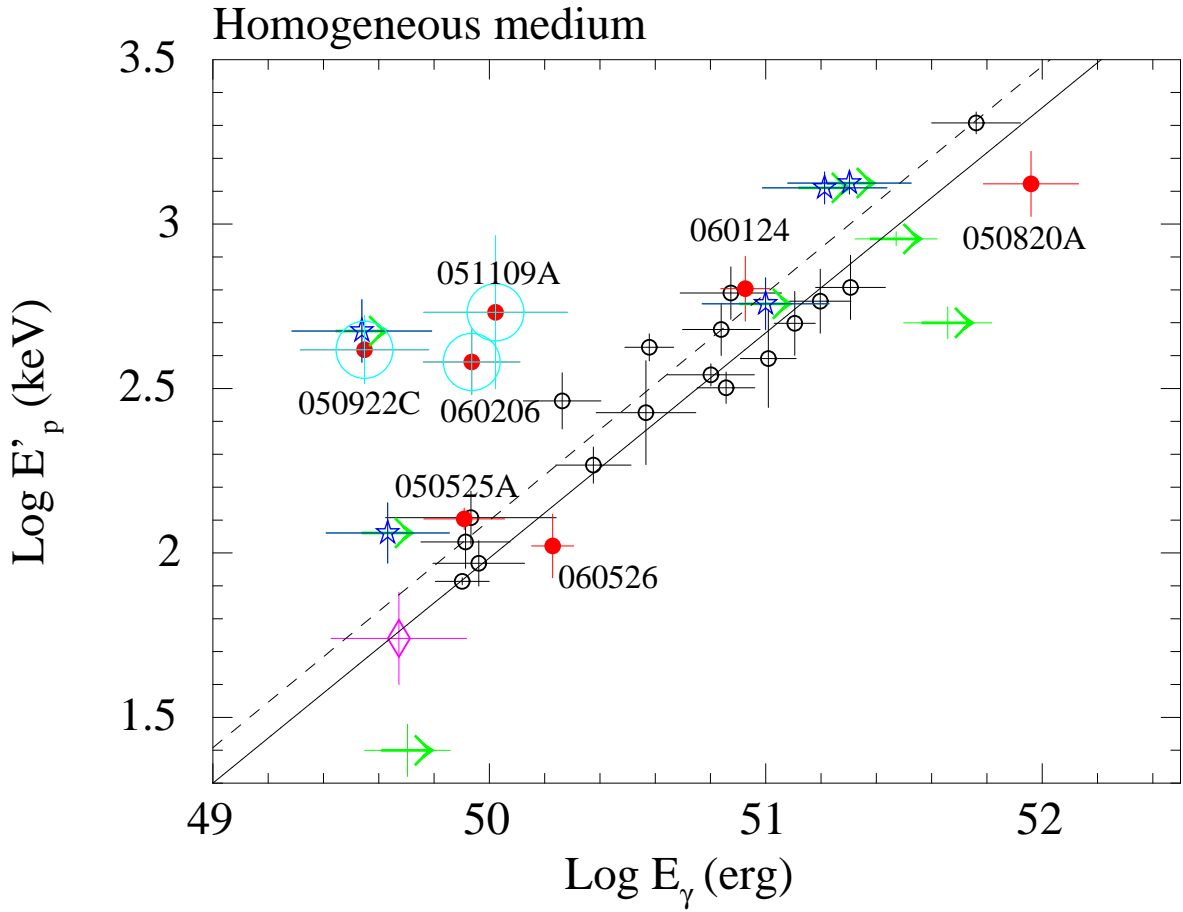}
\end{center}
 
\caption{The rest-frame spectral peak energy $E'_{\rm p}$ and the collimation
corrected energy $E_\gamma$. This energy has been calculated assuming a
homogeneous density medium with a value of $n_0 = 3$~cm$^{-3}$. The open circles
mark the pre-\textit{Swift} bursts used to derive the Ghirlanda relation in
Nava et al. (2006). Filled circles represent \textit{Swift} GRBs
with known peak energy, redshift, and break time in the optical. The arrows mark
\textit{Swift} bursts with only lower limits in the break time $t_{\rm b}$.
Those events marked with stars lie at the left of the best-fit correlation
and are consistent with it, so they have not been included in the fit. 
Events with a big circle around are the two GRBs with the optical break not
coincident with the X--ray break, which have been discarded in the additional
analysis. The point marked with a diamond is GRB\,060614, shown for comparison
purposes.
The dashed line shows the best-fit power law model for the events in the Nava
et al. (2006) sample, obtained by accounting for the errors on both
coordinates. The solid line shows the same fit to the previous sample with
the addition of all \textit{Swift} bursts (except those marked with stars).
\label{fg:homo}} 

\end{figure*}

\begin{figure*}[htbp]
\begin{center}
\psfig{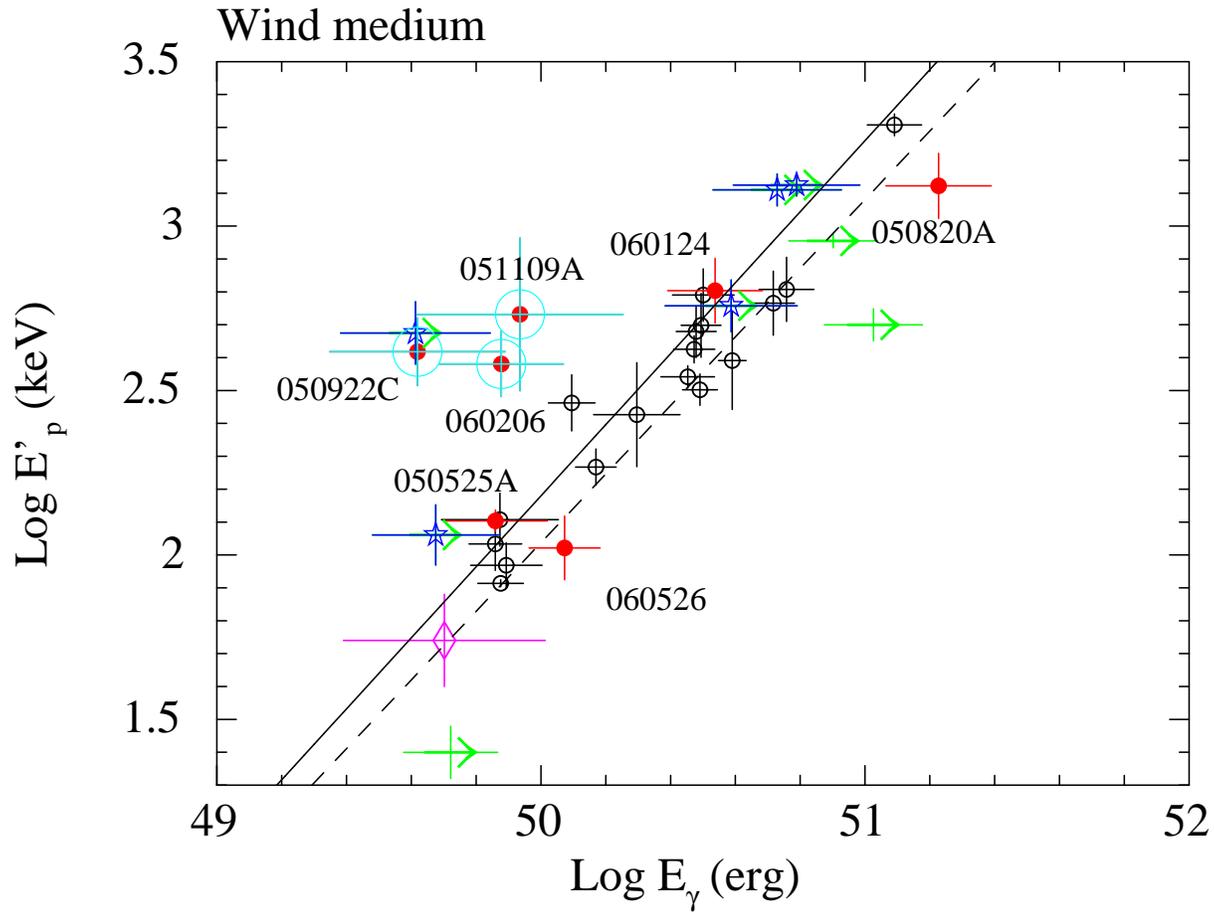}
\end{center}
\caption{As in the previous figure but for a wind medium with density parameter $A_*=1$.
\label{fg:wind}}
\end{figure*}

We added the new \textit{Swift} bursts to the sample of 18 events presented by
Nava et al. (2006). 
We note that GRB\,050525A is common to both samples. There are eight 
\textit{Swift} bursts with a measurement of $E_{\rm iso}$, $E_{\rm p}$ and 
$t_{\rm b}$ (GRB\,060614 is not used however), plus eight events with a
lower limit on $t_{\rm b}$. In Fig.~\ref{fg:homo} and \ref{fg:wind} we
show the \textit{Swift} bursts in the $E'_{\rm p}$ vs. $E_{\rm iso}$ plane.

We repeated the analysis carried out by Nava et al. (2006) on their sample using
the routine {\em fitexy} of Press et al. (2003) and fit the Ghirlanda relation
to their data for both the ISM and wind cases. We obtained consistent results (see
Table 2), adopting the value of $n_0= 3$~cm$^{-3}$ for the circumburst 
density and $\eta_\gamma = 20$\% for the gamma--ray efficiency for all events. 

Before starting the statistical analysis, we observed that five bursts with
lower limits on the break time lie at the left of the best-fit Ghirlanda relation
(GRB\,050318, GRB\,050603, GRB\,060418, GRB\,060927 and GRB\,061121, see
Figs.~\ref{fg:homo} and \ref{fg:wind}). Since they have only lower limits on
the corrected energy and are consistent with the relation, they are not
considered in the fit. 

We were left with 27 GRBs. As can be seen in Figs.~1 and 2, there are 
significant outliers to the relation. Adopting the routine {\it fitexy}, we
derived a reduced $\chi^2 = 3.82$ (25 dof) for the homogeneous ISM 
relation (null hypothesis probability of $4\times 10^{-10}$) and a reduced 
$\chi^2 = 3.09$ (25 dof) for a wind ISM relation (null hypothesis 
probability of $3\times 10^{-7}$). We note that our limits are 
conservative, since we had treated the lower limits on $t_{\rm b}$ as actual
measurements. If the breaks occurred significantly later than these limits,
many of the bursts marked by arrows in  Figs.~\ref{fg:homo} and \ref{fg:wind}
would move rightward in the plot, even more increasing the scatter in the
correlation. 

Given the concerns about using the Ghirlanda relation in terms of the
jet interpretation, we also fitted the model-independent Liang-Zhang relation
to the bursts in the sample. Again, by including only the events listed in the
Nava et al. (2006) sample, we found consistent results with what they found,
albeit with a slightly larger $\chi^2$ (see Table 2). Using the data as
listed in the Tables 1 and 3 of Firmani et al. (2006), we find a 
significantly larger $\chi^2$, which is entirely due to GRB\,050525A (variation 
in the reduced $\chi^2$ from 0.83 to 2.46). We then added the
\textit{Swift} bursts to the sample, and again found that the relation
worsened significantly. Moreover, the exponent $y$ in the Liang-Zhang relation 
is now $y = -0.55 \pm 0.10$ ($68\%$ confidence level),
significantly different from the value $y = -1$ needed to make the Ghirlanda
and Liang-Zhang relations mutually consistent (Nava et al. 2006).

\begin{table*}
\caption{Fit results}
\centering\begin{tabular}{cccccc} \hline\hline
Correlation    & Nava [our] analysis      & 
\textit{Swift} data      & \textit{Swift} data achromatic breaks &
\textit{Swift} data pure breaks\\
               & $\chi^2_{\rm red}$ (dof) & $\chi^2_{\rm red}$ (dof) & $\chi^2_{\rm red}$ (dof) &  $\chi^2_{\rm red}$ (dof)     \\ \hline
Ghirlanda ISM  & 1.40 [1.42] (16)         & 3.82 (25)                & 2.35 (22)                & 2.33 (20) \\
Ghirlanda wind & 1.13 [1.13] (16)         & 3.09 (25)                & 2.00 (22)                & 1.72 (20) \\  
Liang \& Zhang$^*$& 1.49 [1.81] (15)      & 4.56 (24)                & 2.44 (21)                & 2.37 (19) \\ \hline\hline
\end{tabular}

$^*$ The logarithmic errors in this case have been evaluated as the mean of the
lower and higher errors.

\end{table*}

\section{Discussion and conclusions}

We have investigated in detail the addition of \textit{Swift} bursts to the
Ghirlanda and Liang-Zhang relations. Given the narrow energy band of
the BAT instrument, only a small fraction of the $>200$ GRBs so far discovered
by \textit{Swift} precisely determine the peak energy needed for
the comparison with the Ghirlanda correlation. Putting together data from
different instruments (mostly \textit{Wind}-Konus, and occasionally HETE-2,
RHESSI, and \textit{Suzaku}) and including the few bursts for which a prompt
measurement of $E_{\rm p}$ was provided by BAT, we find a total of 19 GRBs. Of
these, eight have a determination of the optical light curve
break (seven are used), eight have a lower limit (three are used), and three
not have enough (published) optical data to constrain it. 

With the inclusion of these nine new GRBs (one was already in the Nava
sample), the Ghirlanda correlation suffers
from a worsening. Fitting the data, the reduced $\chi^2$ goes from 1.1 (1.1) 
to 3.8 (3.1) in the case of a homogeneous (wind) medium (see Table 2).
The correlation is clearly still present, but it is not as tight as suggested in
previous works. Its use for cosmological purposes is thus weakened. Actually,
Spearman's and Kendall's (tau) correlation tests give a relatively weak
probability of correlation on the \textit{Swift} sample (seven objects). With 
these tests we obtain correlation probabilities of $\sim 20\%$ for both
homogeneous and wind cases, to be compared with $10^{-6}$ -- $10^{-8}$ 
for the pre-\textit{Swift} sample.  

In Table 1 we also report the predicted break times based on the Ghirlanda
relation in order to see if there are corresponding breaks in the X--ray light 
curves. On average the predicted breaks do not occur in concordance with X--ray 
breaks. This means that even when
considering breaks in the X--ray light curves, we are not able to reconcile
the observed times with the ones predicted by the Ghirlanda relation. 

The optical light curves of bursts discovered by \textit{Swift} are
significantly more complex than previously thought (e.g. Stanek et al. 2006;
Gal-Yam et al. 2006). One might therefore speculate that the breaks observed in
some of the strong outliers are different from those
responsible for the existence of the Ghirlanda correlation. For example,
GRB\,050922C and GRB\,060206 had breaks both in the optical and in the X--ray
light curves, but at different times (Panaitescu et al. 2006; Stanek et al.
2006). The use of these breaks (so-called chromatic breaks) for the Ghirlanda
relation is not appropriate in principle, since a jet break produces an
achromatic steepening. Only GRB\,060124, GRB\,050525A and GRB\,060614 show
evidence for an achromatic break at a known redshift. Recently,
several papers suggest a different origin of the X--ray and optical
emission (Uhm \& Belobodorov 2007; Grenet, Daigne \& Mochkovitch 2007;
Ghisellini et al. 2007), so that the need for achromatic breaks might be relaxed.

Optical breaks not coincident with X--ray breaks might have been missed
before the launch of \textit{Swift} due to the lack of X--ray coverage. For
the outliers lying above (below) the correlation, a further break at later
(earlier) times (the true ``jet-break'') 
would be necessary in order to reconcile them with the Ghirlanda relation. This
does not always look like a viable solution. For example, considering outliers in
the upper part of the $E'_{\rm p}$--$E_\gamma$ plane, GRB\,060206 showed a steep
decay ($\alpha \approx 2$) after $t_{\rm b} = 0.6$~d (Monfardini et al. 2006;
Stanek et al. 2006), similar to many of the breaks used to build the original
Ghirlanda relation. Even if no observations are available at late times, a
further break appears unlikely. Late observations were probably secured (but
not published) for GRB\,050922C, which would be particularly interesting in
this respect. Concerning the bursts lying on the right side of the
correlation, early observations are available for GRB\,050401 (Rykoff et al.
2005; Watson et al. 2006), which shows a remarkable, unbroken decay starting
$\approx 35$~s after the burst. Thus, at least for this burst, we have strong
``evidence for absence'' of a break that might have reconciled GRB\,050401
with the Ghirlanda relation.

In addition, considering only bursts with achromatic breaks and upper
limits (i.e. we disregard GRB\,050922C, GRB\,051109A and GRB\,060206; we retain
GRB\,050820A since we do not have information from the X--rays), the
results of the fits are improved, with a reduced $\chi^2$ of 2.35 in the case 
of a homogeneous medium (null hypothesis probability of $4\times 10^{-4}$) and
with a reduced $\chi^2=2.00$ (null hypothesis probability of $0.003$) in the
case of a wind medium. As described above, concerns have also been expressed 
on GRB\,050525A and GRB\,060124 since their post break decay index is too flat
and the decay variation is less than one. We thus also consider the case of only
GRB\,050820A and GRB\,060526 and the upper limits (pure breaks in Table 2). 
In this case we obtain a reduced $\chi^2$ of 2.33 for a homogeneous medium (null 
hypothesis probability of $7\times10^{-4}$) and $\chi^2=1.72$ (null hypothesis
probability of $0.02$) for a wind medium.

One might speculate on whether the \textit{Swift} bursts have peculiar properties.
In this respect, it is interesting to note that neither GRB\,050922C nor
GRB\,050603 (the two strongest outliers of the relation) have any particularity,
in terms of their duration, spectral properties, and fluence in the
\textit{Swift} sample. A distinguishing property of \textit{Swift} bursts as
a sample is their average measured higher redshift (e.g. Jakobsson et al. 2006). The
average redshift of the \textit{Swift} burst introduced in our sample is
$\langle z \rangle = 2.2$, to be compared with $\langle z \rangle = 1.4$ for
the events in the sample of Nava et al. (2006). This might hint at a possible
evolutionary effect, possibly due to the the properties of the jet propagation
inside the progenitor stars and then in their environment during the afterglow
phase. The current sample is, however, still too limited to draw any strong
conclusion.  

\begin{acknowledgements}
This work is supported by ASI grant I/R/039/04 and MIUR grant 2005025417. 
We thank S. Barthelmy and J. Greiner for maintaining their GRB web
pages. We also thank D. Malesani and A. Kann for detailed discussions 
as well as G. Ghirlanda et al. (2007, astro-ph/0704.0234) for comments. 
This work has made use of the GRBlog database. 
\end{acknowledgements}

\end{document}